# Reduction of laser intensity scintillations in turbulent atmospheres using time averaging of a partially coherent beam

G P Berman[1], A R Bishop[2], B M Chernobrod[1], V N Gorshkov[1,3,4], D C Lizon[5], D I Moody[6], D C Nguyen[5], S V Torous[3]

[1]Theoretical Division, T-4 & CNLS, MS B213, Los Alamos National Laboratory, Los Alamos, New Mexico 87545
[2]Theory, Simulation & Computation Directorate, MS B210, Los Alamos National Laboratory, Los Alamos, New Mexico 87545
[3]National Technical University of Ukraine "KPI," 37 Peremogy Avenue, Building 7, Kiev-56, 03056 Ukraine
[4]The Institute of Physics, National Academy of Sciences of Ukraine, Nauky Ave. 46, Kiev 680028, Ukraine
[5]ISR-6, MS H851, Los Alamos National Laboratory, Los Alamos, New Mexico 87545
[6]ISR-2, MS B244, Los Alamos National Laboratory, Los Alamos, New Mexico 87545

E-mail: **gpb@lanl.gov**

Abstract

We demonstrate experimentally and numerically that the application of a partially coherent beam (PCB) in combination with time averaging leads to a significant reduction in the scintillation index. We use a simplified experimental approach in which the atmospheric turbulence is simulated by a phase diffuser. The role of the speckle size, the amplitude of the phase modulation, and the strength of the atmospheric turbulence are examined. We obtain good agreement between our numerical simulations and our experimental results. This study provides a useful foundation for future applications of PCB-based methods of scintillation reduction in physical atmospheres.

*PASC:* 42.25.Dd

1. Introduction

Intensity scintillations due to the influence of atmospheric turbulence are a major obstacle for gigabit data rates and long-distance, free-space optical communications (FSOC) [1,2]. Several approaches have been developed to greatly reduce turbulence effects, including adaptive optics, aperture averaging, array receivers, and PCBs (beams with multiple coherent spots in their transverse section). See the detailed reviews [1,3]. Nevertheless, scintillations continue to limit the information capacity of FSOC channels. It is well-known that PCBs are less affected during propagation through atmospheric turbulence than a fully coherent beam [4-13]. Recently it was shown [10] that a PCB in combination with time averaging results in a significant reduction of scintillations. However, this time-averaging (slow detector) technique is incompatible with a gigabit data rates. Recently we proposed a new method for scintillation reduction, combining time averaging of a PCB with spectral encoding [14]. In our approach, information is encoded in the form of dips in the spectrum of sub-picoseconds coherent laser pulses. During a multipulse sequence with the same encoded information, the partially coherent pattern is changed several times. Before the receiver plane, the laser beam passes through a grating. After the grating each spectral bit is directed to a certain angle and is received by a certain pixel in a linear CCD array. The relatively slow CCD averages these partially coherent patterns in time. This time averaging results in a significant reduction of the scintillation index. Our proposed method provides a gigabit data rate channel with a reduced scintillation index.

In this paper, we present the results of our experimental and numerical study of scintillation reduction using time averaging of a PCB. The main goal of this paper is the optimization of the essential parameters, including the coherence radius of the PCB, the frame rate of spatial light modulator (SLM), and the spatial resolution and frame rate of the CCD sensor. Our experimental approach is very similar to one used in [10]. In [10], the authors used a simplified approach, in which a rotating phase diffuser simulated the atmospheric turbulence, the PCB being simulated by a similar phase diffuser. The experiments [10] were done with the pseudo-PCB, in which the laser beam had only a few speckles. In our experiments, we use a PCB with a variable number of speckles and several sizes of signal integration areas. The PCB demonstrates how the scintillation reduction efficiency depends on the number of speckles. Our numerical simulations include a variety of phase diffusers and control parameters, including a phase modulation index, an inhomogeneity radius, and radius dispersion. We also examine the influence of the strength of the atmospheric turbulence on the efficiency of scintillation reduction.

## II. Numerical simulations

In our model, a PCB propagates along the $z$-axis and has a phase distribution, $\phi_{z=0} = \phi(x,y,t)$, where $\phi(x,y,t)$ is a random function with a defined coherence radius and dispersion. The time-dependent phase modulation can be written as function of the random parameters, $c_k(t)$, $k=1,2,...,K$; $\phi_{z=0} = \varphi(x,y;c_1(t),c_2(t),...,c_K(t))$. For example, if a laser beam with radius, $r_0$, passes through a diffuser with radius, $R_{Dif} \gg r_0$, rotating in the $x$-$y$-plane, then $\phi_{z=0} = \varphi(x-c_1(t), y-c_2(t))$, where $\varphi(x,y)$ is a deterministic function. We denote all of the control random parameters by $\{C(t)\}$. Since the atmospheric turbulence is relatively slow process, the refractive index is a slow function of time. Thus, for the short time interval $\tau_0(l-0.5) < t < \tau_0(l+0.5)$, we assume that $n(x,y,z,t) \approx n(x,y,z,l\tau_0) \equiv n^l(x,y,z)$, where $l$ is the number of time interval with duration $\tau_0$. The integral light intensity, $I^l$, corresponding to this time interval is given by $I^l(S) = \int_{(S)} I_r^l(x,y,z=D)ds$, where $S$ is the area of the photosensor, $I_r^l(x,y,z=D)$ is the light intensity distribution at the photosensor plane, and $D$ is the distance between the transmitter and receiver. The scintillation index is given by

$$\sigma^2 = \frac{\langle (I^l)^2 \rangle - \langle I^l \rangle^2}{\langle I^l \rangle^2}, \tag{1}$$

where the averaging indicates a sum over the indexes, $l$, assuming that the averaging time is infinite. This corresponds to averaging over all atmospheric statistical states. It is to be noted that $\sigma^2$ depends on the photosensor area, $S$, namely, the larger the area, the smaller scintillation index.

Usually, the value of scintillation index, $\sigma^2$, limits the quality of the FSOC. To reduce the scintillation index, the PCB can be applied in combination with time averaging:

$$\hat{I}^l(S) = \frac{1}{\tau_0} \int_{(S,\tau_0)} I_r^l[C(t); x, y, z = D] ds dt. \tag{2}$$

For a rather fast SLM with the frame rate equal to $1/\tau_{SLM}$ ($\tau_{SML} \ll \tau_0$), a complete stochastic sequence of the control parameters, $\{C\}$, can be achieved and the reduction of the scintillation index can be very significant

$$\hat{\sigma}^2 \ll \sigma^2, \tag{3}$$

where $\hat{\sigma}^2$ is the scintillation index obtained for the sequence $\{\hat{I}^l(S)\}$. However, for a Gigabit channel, the frame rate of the fastest SLM is not adequate to fulfill the condition $\tau_{SLM} \ll \tau_0$.

As we mentioned in the Introduction, information is encoded in the spectrum modulation of the laser pulse, every pulse with duration of $\tau_0$ transmitting $N$ bits. In this case, the same information ($N$ bits) can be transmitted $M$ times as the sequence of repeated pulses with the total time interval, $\tau = M(\tau_0 + \tau_1) \gg \tau_0$, where $\tau_1$ is the time duration between neighboring pulses. During the time interval with the sequence number $m$, $(m-1)\tau < t < m\tau$, the SLM generates the phase distribution function, $\varphi(x, y; C(t))$. An individual bit has the signal

$$\hat{I}_m^l(\Delta\omega) = \int_{(S)} I_r^l(C(t_m); x, y, z = D, \Delta\omega) ds, \tag{4}$$

where $\Delta\omega$ is the frequency interval that corresponds to the bit, $t_m = (m - 1/2)\tau$.

Since the information is repeated $M$ times, the following average signal corresponds to each bit:

$$\hat{I}^l = \frac{1}{M} \sum_{m=1}^{M} \hat{I}_m^l. \tag{5}$$

(The parameter $\Delta\omega$ is omitted to simplify subsequent expressions).

In order to decrease the scintillation index, $\hat{\sigma}^2$, the condition, $\tau_{SLM} \ll \tau$, should be satisfied. Then, we generate $M$ statistically independent realizations of the SLM phase distribution, $\varphi(x, y; C(t_m)) \equiv \varphi_m(x, y), m = 1, 2, ..., M$. Accordingly, the signals $\hat{I}_m^l$ in (4) are statistically independent, too. In this case

$$\hat{\sigma}^2 = \frac{\sigma_s^2}{M}, \tag{6}$$

where $\sigma_s^2$ is the scintillation index of the individual signals $\hat{I}_m^l$:

$$\sigma_s^2 = \left(\left\langle(\hat{I}_m^l)^2\right\rangle - \left\langle\hat{I}_m^l\right\rangle^2\right)\Big/\left\langle\hat{I}_m^l\right\rangle^2. \tag{7}$$

Averaging is performed over both indexes, $m$ and $l$.

Thus, for the effective suppression of scintillations of the averaged signals (5), two conditions should be fulfilled:

**A.** $\sigma_s^2 < \sigma^2$. (Recall that $\sigma^2$ corresponds to individual signals passed only through the turbulent atmosphere; $\sigma_s^2$ corresponds to individual signals passed through both a SLM and the turbulent atmosphere.) Whether this inequality is met depends on the properties of the SLM used.

(8)

**B**. For a given index, $l$, the signals, $\hat{I}_m^l$, have to be statistically independent. Otherwise the effectiveness of the averaging (5) diminishes: $\hat{\sigma}^2$ in (6) is not proportional to $1/M$.

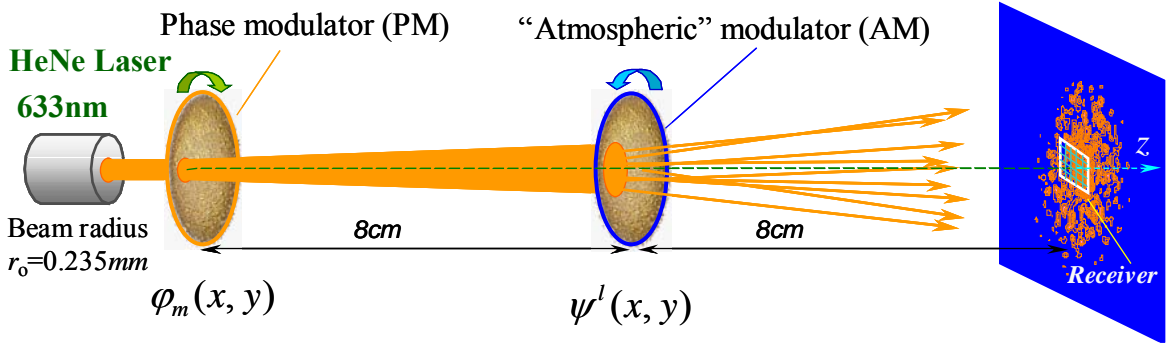

**Figure 1**. Schematic diagram of the setup for numerical study of scintillations reduction by using time averaging of PCB. $\varphi_m(x, y)$ is the phase distribution at the plane $z = 0$ introduced by the phase modulator. $\psi^l(x, y)$ is the phase distribution introduced by the "atmospheric" modulator.

*Mathematical model.*. Numerical simulations were performed for the model shown in Figure 1, where the PCB is created by a phase modulator. The atmospheric turbulence is simulated by the SLM (atmospheric modulator). We will investigate in the paraxial approximation the spatial evolution of a stationary linearly polarized light beam with vector potential, $\mathbf{A} = \mathbf{e}U(x, y, z)\exp[i(\omega t - kz)]$, (where ω is the frequency and $k = 2\pi/\lambda$ is the wave vector). The complex amplitude, $U(x,y,z)$, satisfies the Leontovich parabolic equation,

$$\frac{\partial U}{\partial \tilde{z}} = \frac{1}{4i}\left[\frac{\partial^2 U}{\partial \tilde{x}^2} + \frac{\partial^2 U}{\partial \tilde{y}^2}\right], \qquad (9)$$

where $e$ is the unit vector in the direction of light polarization. The values of the magnetic and electric fields, $B$ and $E$, are given by $B = rotA$ and $E = -\frac{1}{c}\frac{\partial A}{\partial t}$. The light intensity is $I(x,y,z) \sim |U(x,y,z)|^2$. We use the dimensionless coordinates $\tilde{z} = z/z_R$, $\tilde{x} = x/r_0$, $\tilde{y} = y/r_0$, where $z_R = \frac{\pi r_0^2}{\lambda}$ is the Rayleigh length, $r_0 = 0.235\ mm$ is the beam radius, $\lambda = 0.633\mu m$ is the wavelength (the dimensionless beam radius $\tilde{r}_0 = 1$). The optical field in front of the phase modulator (PM) has the Gaussian distribution, $U(x,y,z=0) = U_0\exp(-r^2/r_0^2)$, and after PM it becomes the PCB, $U(x,y,z=+0) = U(x,y,z=0)\exp[i\varphi_m(x,y)]$.

Our numerical calculations are performed in the interval $0 < z < 16\ cm$. At the distance, $z = 8cm$, after passing through the atmospheric modulator (AM), the field acquires an additional phase factor $\exp\left[i\psi^l(x,y)\right]$. The equation (9) is solved numerically by the difference method, the Pismen-Reckford scheme being employed. Our FORTRAN code uses the IMSL library. The finite-difference grid has uniform spacings:

$$\Delta\tilde{x} = \Delta\tilde{y} = \tilde{r}_0/100 = 0.01,\ 0 \leq \tilde{x} \leq N\Delta\tilde{x},\ 0 \leq \tilde{y} \leq N\Delta\tilde{y},\ N = 2^{12} = 4096.$$

At the boundary of the computational space, we set $U(\tilde{x}, \tilde{y}, \tilde{z}) = 0$. We do not use a Fourier expansion (FFT subroutine in the IMSL library) because this method has significant errors for complicated fields and a large number of nodes. The signals, $\hat{I}_m^l$, are calculated for three CCD integration areas with the centers on the z-axis with size: $R1$ – 1 pixel, $R2$ - $15 \times 15$ pixels, and $R3$ - $20 \times 20$ pixels. The size of each individual pixel is $2.8 \times 10^{-3} \times 2.8 \times 10^{-3}\ mm^2$. The random functions, $\psi^l(x,y)$ and $\varphi_m(x,y)$, are generated by using the same algorithm with different stochastic parameters $\Lambda_0, \gamma, A_0$ (see below). The random function is:

$$\varphi(x,y) = \text{Re}[\zeta(x,y)],$$
$$\zeta(x,y) = \sum_{n=1}^{N_x}\sum_{m=1}^{N_y}[a(n,m) + ib(n,m)] \times \exp\left[i\left(xk_n^{(x)} + yk_m^{(y)}\right)\right] \qquad (10)$$
$$k_n^{(x)} = \frac{2\pi}{L}n,\ k_m^{(y)} = \frac{2\pi}{L}m$$

where $L = 4096 \cdot \Delta x \approx 9.6 mm$. The random parameters, $a(n,m)$, $b(n,m)$, are defined by

$$k_{nm} = \sqrt{\left(k_n^{(x)}\right)^2 + \left(k_m^{(y)}\right)^2}, \ \Lambda = 2\pi / k_{nm}, \ A_{nm} = A_0 \exp\left[-\left(\frac{\Lambda - \Lambda_0}{\Lambda_0}\right)^2 \times \frac{\gamma^2}{2}\right], \quad (11)$$

$$a(n,m) = A_{nm} \times r^{(a)}, \ b(n,m) = A_{nm} \times r^{(b)}. \quad (12)$$

The random numbers, $r_a$ and $r_b$, have normal probability distributions: $p(r_a) = \frac{1}{\sqrt{2\pi}} \exp(-r_a^2 / 2)$, $p(r_b) = \frac{1}{\sqrt{2\pi}} \exp(-r_b^2 / 2)$. If $|r_{a,b}| > 3.5$, then the generation is repeated to satisfy the inequality, $|r_{a,b}| \leq 3.5$.

We use the following sets of phase modulator parameters:

Modulator 1: $\Lambda_0 = 0.07 mm$, $\gamma = 7$, $A_0 = 0.04$;
Modulator 2: $\Lambda_0 = 0.07 mm$, $\gamma = 7$, $A_0 = 0.03$;
Modulator 3: $\Lambda_0 = 0.08 mm$, $\gamma = 5$, $A_0 = 0.01$.

We use all these modulators in two different ways: as a PM in some numerical simulations, and as an AM in other simulations. Note that, for given beam parameters, the Rayleigh length is $z_R = \frac{\pi r_0^2}{\lambda} \approx 27.5 cm$. The beam radius at the distance $d = 8 cm$ increases by factor $\sqrt{1 + (d/z_R)^2} \approx 1.04$. As a result, the optical field of the coherent beam at the distances $0 < z \leq d$ is almost the same as at the beginning of propagation, $U(x, y, z) \approx U(x, y, z = 0)$. Thus, we can utilize a certain property of the beam, which is useful for analyzing the obtained numerical results:

If a phase modulator is placed at an arbitrary plane, $z = z_{ph}$ ($0 < z_{ph} < d$), the statistical characteristics of the optical field, $U(x, y, z > z_{ph})$, of the laser beam which passed through the modulator depends only on the distance from the phase modulator, $\Delta z = z - z_{ph}$. (13)

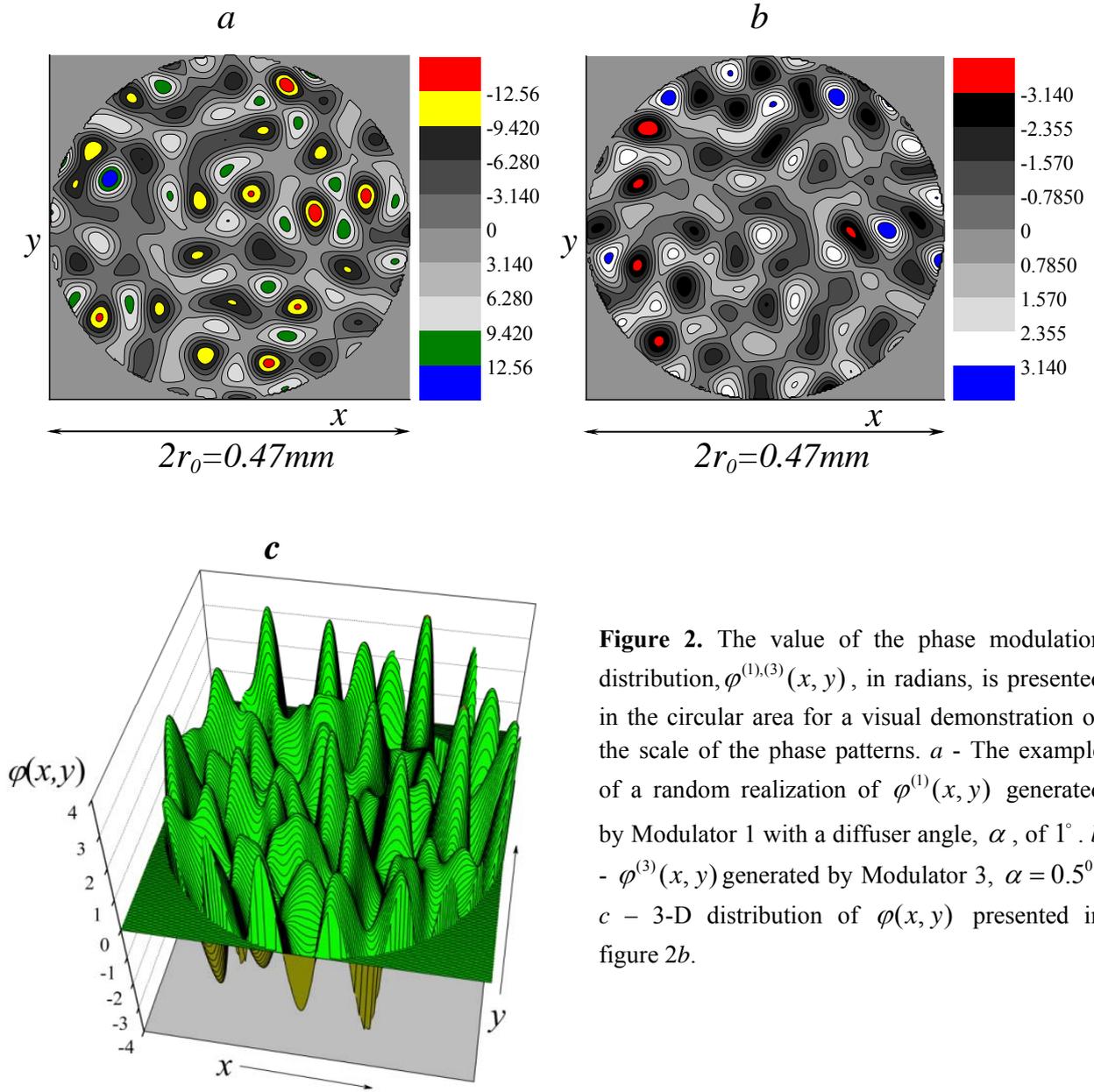

**Figure 2.** The value of the phase modulation distribution, $\varphi^{(1),(3)}(x, y)$, in radians, is presented in the circular area for a visual demonstration of the scale of the phase patterns. *a* - The example of a random realization of $\varphi^{(1)}(x, y)$ generated by Modulator 1 with a diffuser angle, $\alpha$, of $1°$. *b* - $\varphi^{(3)}(x, y)$ generated by Modulator 3, $\alpha = 0.5^0$. *c* – 3-D distribution of $\varphi(x, y)$ presented in figure 2*b*.

*Numerical results.* Figure 2 shows the random realizations of the phase modulation calculated according to Eqs. (10-12) for Modulator 1 (the strongest diffuser) and for Modulator 3. The coherence radius of both modulators is about $r_0/6 \approx 40 \mu m$. This size corresponds to the size of the microlenses covering the surface of diffusers used in our experiments (see below). Figure 3 demonstrates the typical distributions of the intensity of the coherent laser beam after passing through the atmospheric modulator. For the atmospheric modulator, we used Modulator 2 and Modulator 3.

Let us analyze the spatial evolution of coherent beam after the AM (Modulator 3). Figure 4 presents the dependence of the scintillation index on the propagation distance for several signal integration areas of the CCD. Our simulations show that the larger the integration area, the lower the scintillation index.

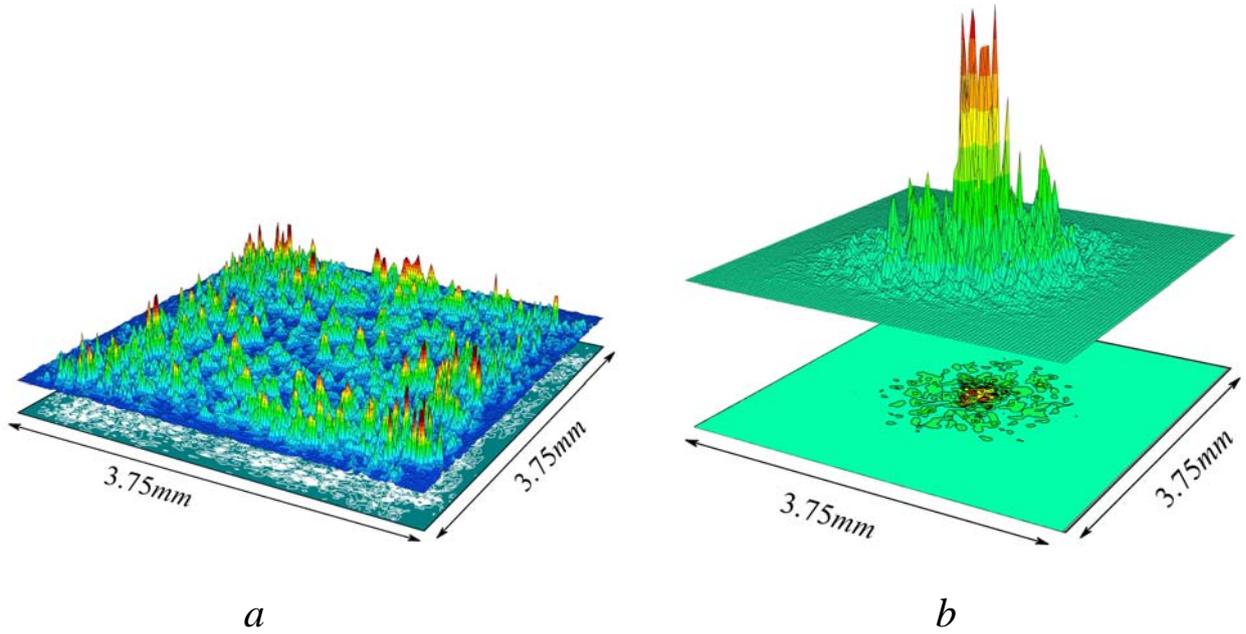

**Figure 3**. The typical distributions of the intensity of coherent laser beam having passed through the atmospheric modulator. Atmosphere is presented by Modulator 1 (*a*) and by Modulator 3 (*b*).

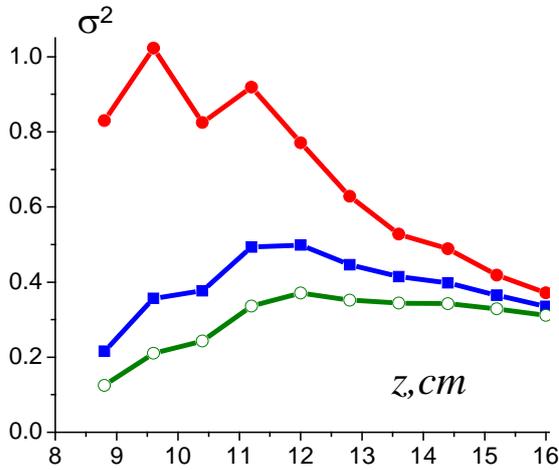

**Figure 4.** The dependence of the scintillation index on the integration area on the CCD. The red, blue, and green curves are associated with areas R1, R2, and R3, correspondingly. The averaging is performed over 1000 realizations of the phase, $\psi^l(x, y)$, $l = 1, 2, ..., 1000$.

In the near-field zone, after the AM ($z \sim 9 cm$), the light beam is fragmented significantly (Figure 5a). Bright zones are separated by many spiral phase dislocations (optical vortices) with zero intensities at the centers. (In Figure 5b the green circle indicates one of the vortices.) For signal integration area, R1 (one pixel size), a bright spot or an optical vortex can cover the whole area. This results in a large value of the scintillation index ($\sigma^2 \sim 1$). The size of the CCD areas, R2 and R3, are comparable with typical sizes of the inhomogeneities. For these conditions the received signal has rather small scintillations. The separation of the fragments increases with increasing distance, $z \sim 10 - 12 cm$. This tendency results in an increase of the scintillation index for the integration areas, R2 and R3. At the same time, the size of fragments increases with distance due to diffraction. This last tendency leads to a decrease of the scintillation index for the small integration area, R1.

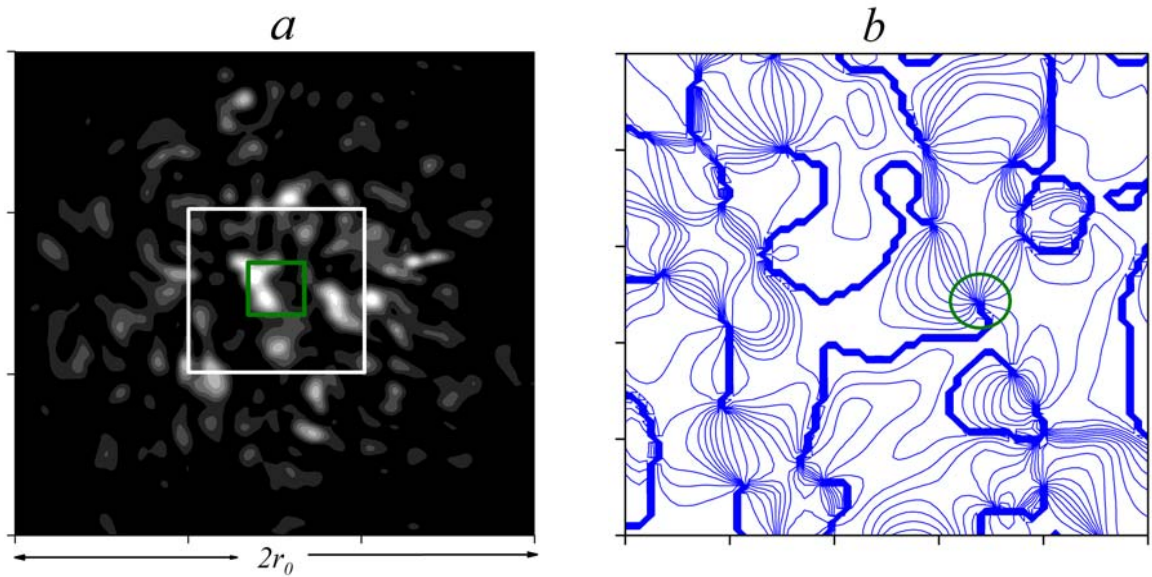

**Figure 5**. *a* -A typical distribution of the intensity at the distance $z = 8.8\ cm$, after passing the atmospheric modulator. The size of the numerical polygon (the total size of the presented 2-D distribution) is $2r_0$. The green square is the signal integration area R3. *b* – The phase distribution corresponding to the area inside the white square in Figure 5a. The green circle in Figure 5*b* indicates an optical vortex with zero amplitude and undefined phase.

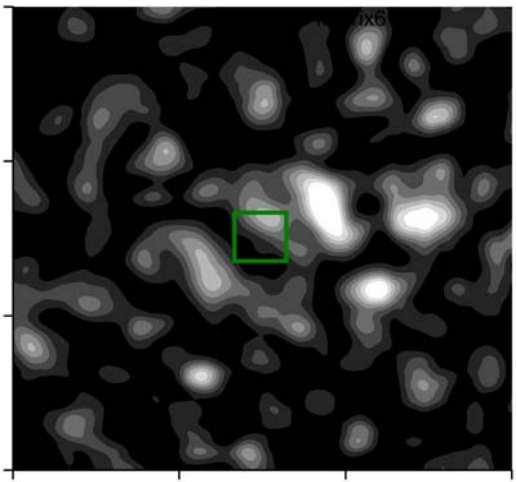

**Figure 6.** Intensity distribution at the distance $z = 12 cm$. The same areas are represented as shown in Figure 5.

In the far field zone, ($z > 12 cm$) individual fragments of the light beam overlap, and the singularities of the wave front (optical vortexes) move to the beam periphery (see Figure 7). (This spatial behavior of the optical vortices is typical when the vortices with different topological charges $(\pm 1)$ appear as a result of small-scale inhomogeneities inserted in the beam by different amplitude-phase masks [15].) As a result, the scintillation index decreases with increasing distance for all sizes of CCD integration areas, and asymptotically tends to $\sigma^2(D) \approx 0.35$.

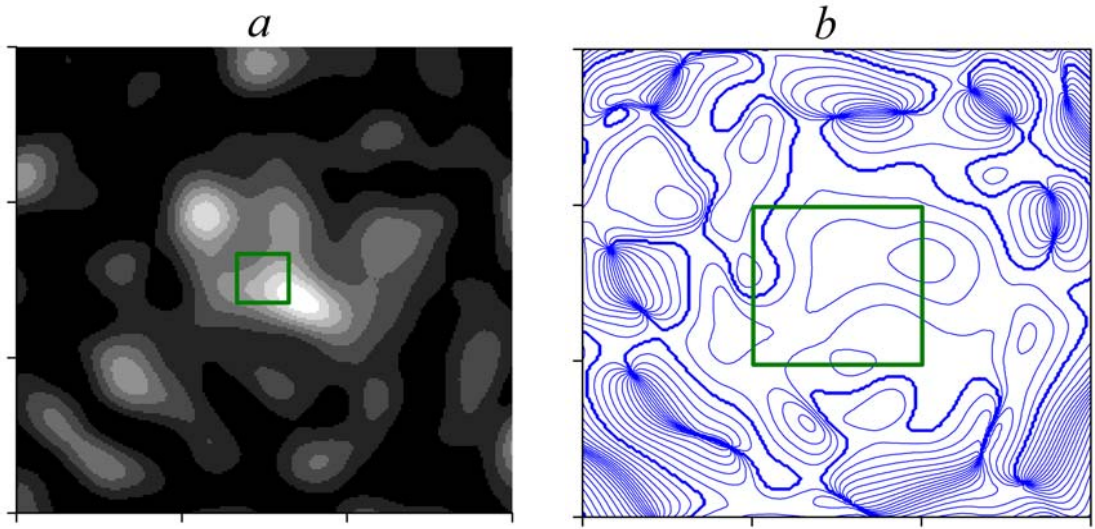

**Figure 7.** Distribution of the intensity (*a*) and phase (*b*) at the distance $z = D$ =16 cm. The size of the 2-D distribution is $2r_0$. The green square in Figure 7*b* corresponds to the green square in Figure 7*a*.

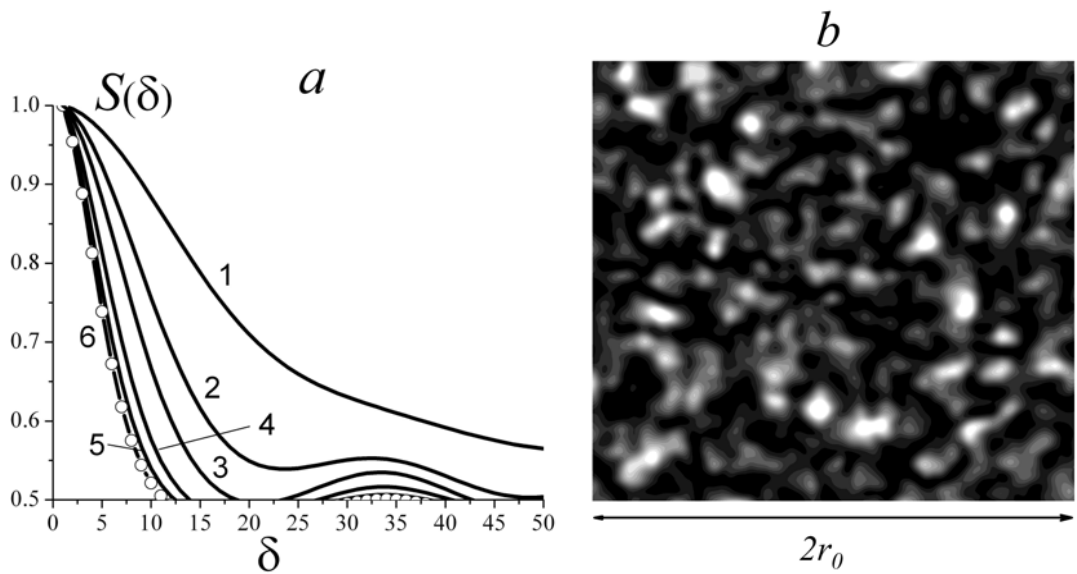

**Figure 8.** *a* - Dependence of the correlation parameter on the distance between two points on the phase front, calculated for the values of coherence beam radius 1 - $R_0 = r_0$, 2 - $R_0 = 3r_0/2$, 3 - $R_0 = 2r_0$, 4 - $R_0 = 3r_0$, 5 - $R_0 = 4r_0$, 6 - $R_0 = 5r_0$. Unit of $\delta$ is $r_0/100$. The AM is Modulator 3. *b* – The intensity distribution $I^1(x,y,D)$ for a broad coherent beam with radius, $R_0 = 4r_0 = 0.94mm$. Compare this distribution with the distribution shown in Fig. 7*a* ($R_0 = r_0$).

***Application of PCB***. We first introduce a few definitions. $I(x,y,z) = I_m(x,y,z)$ is the light intensity in the distance interval $0 \leq z < d = 8cm$ and $m$ is the index of the PM realization. For the PM we used Modulator 2. For the distances $d \leq z \leq D = 16cm$, we introduce

$$I(x,y,z) = \begin{cases} I^l(x,y,z) \text{ for coherent laser beam,} \\ I_m^l(x,y,z) \text{ for PCB.} \end{cases}$$

The index, $l$, is the index of the AM realization.

To satisfy (8A), $\sigma_s^2 < \sigma^2$, a small spatial scale for the speckle structure at the plane of the CCD is preferable. In this case, many speckle spots of varying intensity distribution can cover the integrating CCD area. As a result, the integral intensity can have small fluctuations. One of the possible solutions is to use a coherent beam with a large radius, $R_0$. Figure 8 shows the dependence of correlation parameters

$$S_x(\delta) = \frac{\langle I^l(x,y,D) \times I^l(x+\delta,y,D) \rangle}{\langle [I^l(x,y,D)]^2 \rangle}, \quad S_y(\delta) = \frac{\langle I^l(x,y,D) \times I^l(x,y+\delta,D) \rangle}{\langle [I^l(x,y,D)]^2 \rangle} \quad (14)$$

on the beam radius. (Modulator 3 is used as the AM.) In our case, $S_x(\delta) \approx S_y(\delta) = S(\delta)$. These simulations show that, for a coherent beam with a beam radius at the CCD plane equal to $5r_0$, the correlation parameter is $\sim 0.075 r_0$.

However, this case has a disadvantage: the coherent laser beam does not provide the time averaging (5) over different states, $m$, because the phase distribution at the plane of the transmitter does not change in time. The application of a widened PCB is much preferable. To demonstrate the utility of PCBs, we discuss below the results of numerical experiments with several sets of the control parameters of the PM and AM modulators. Note that both of the factors (i) the small spatial scale of the speckle structure of PCB, and (ii) statistical independence of different distributions, $I_m(x,y,z)$, lead to a decreased scintillation index, $\hat{\sigma}^2$. The contributions of these two factors depend on the parameters of AM and PM.

***Numerical simulations 1.*** Modulator 3 serves as the atmospheric modulator. For preparation of the PBC, we use Modulator 2. The results of our numerical simulations are shown in Figure 9. The scintillation index is calculated for all three signal integrating areas, R1, R2, and R3. At the AM plane the beam radius $R(d) \approx 5.5 r_0$. After the AM, the beam fragmentation does not change: one can see (Fig. 9b) that the index, $\sigma_s^2$, calculated for the integration area, R1, is about a constant. According to the data in Figure 9b, $\hat{\sigma}^2 \approx \sigma_s^2 / 10$. Consequently, signals with different values of $m$ are statistically independent (See Exp. 6; the condition (8B) is satisfied.) However, the desirable condition (8A), $\sigma_s^2$ (in Fig. 9b) < $\sigma^2$ (in Fig.4), is not met. Nevertheless, the scintillation index is reduced (mostly due to averaging (5)) by factor of 8 compared with the coherent laser beam.

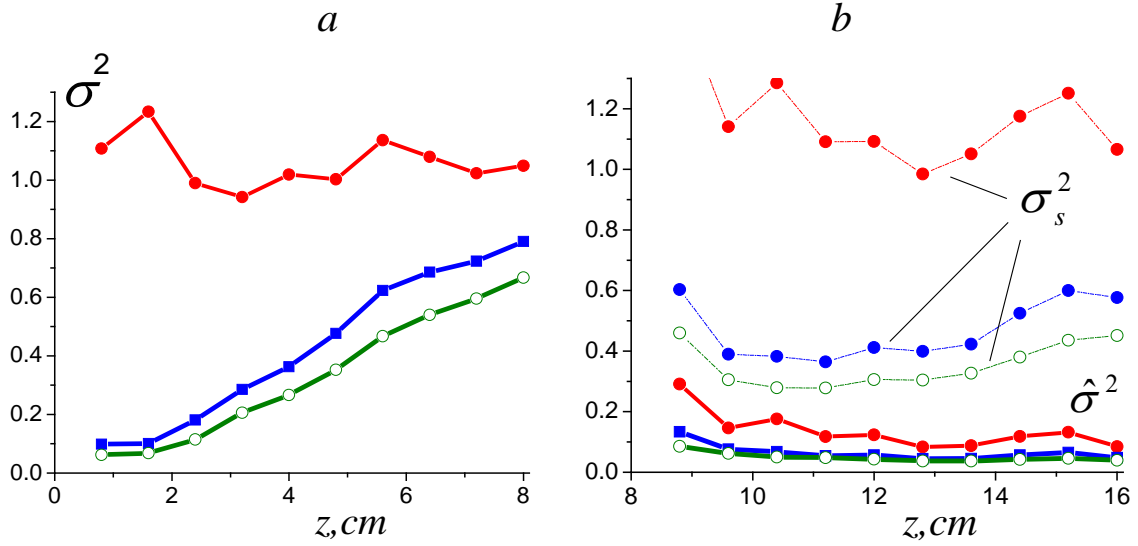

**Figure 9.** The results of the numerical simulations 1: PM – Modulator 2, AM – Modulator 3. *a* – The scintillation index after the first modulator PM. Averaging is made over 1000 PM states. The different colors of the curves correspond to the different signal integration areas. The corresponding colors are the same as in Figure 4. *b* – The scintillation indexes after both modulators, PM and AM. When calculating $\hat{\sigma}^2$, averaging (5) is realized at $M = 10$ for 100 states of AM.

*Numerical simulations 2.* To enhance the scintillation reduction, we choose Modulator 1 as the PM. This modulator has higher modulation index compared with Modulator 2 (see Figure 2a). The speckle size at $z < d$ becomes smaller in comparison with the preceding numerical experiment. When comparing the results shown in Fig. 9*a* and Fig. 10*a*, it can be observed that for the signal integration areas, R2 and R3, the scintillation index is smaller by a factor of 2. In addition, at the plane of the AM, the PCB radius is larger, $R(d)/r_0 \approx 7$. (In Figure 9, $R(d)/r_0 \approx 5.5$.) An example of the intensity distribution at $z = d$ is presented in Figure 3a.

The additional spreading of the PCB results in a decrease in the scintillation index, $\sigma_s^2$, compared with the coherent beam (at $12 cm < z < D$ the ratio, $\sigma^2/\sigma_s^2$, varies from 1.5 to 1.3). The averaging of the signal $I_m^l$ over 10 realizations of the PM phase distribution, $\varphi_m(x, y)$, improves the scintillation index by the factor 1/10 with high accuracy. As a result, the averaging over PCB realizations gives a reduction of the scintillation index by factor of ~14. (According to the results in Figures 4,10c, $\sigma^2 \approx 0.32$ and $\hat{\sigma}^2 \approx 0.023$ for the integrating area R3 at $z = 16 cm$). Note that the numerical simulations 2 correspond to our actual experiments, which are described below, with rather high accuracy. The reduction factor of the scintillation index obtained in our experiments is 16.

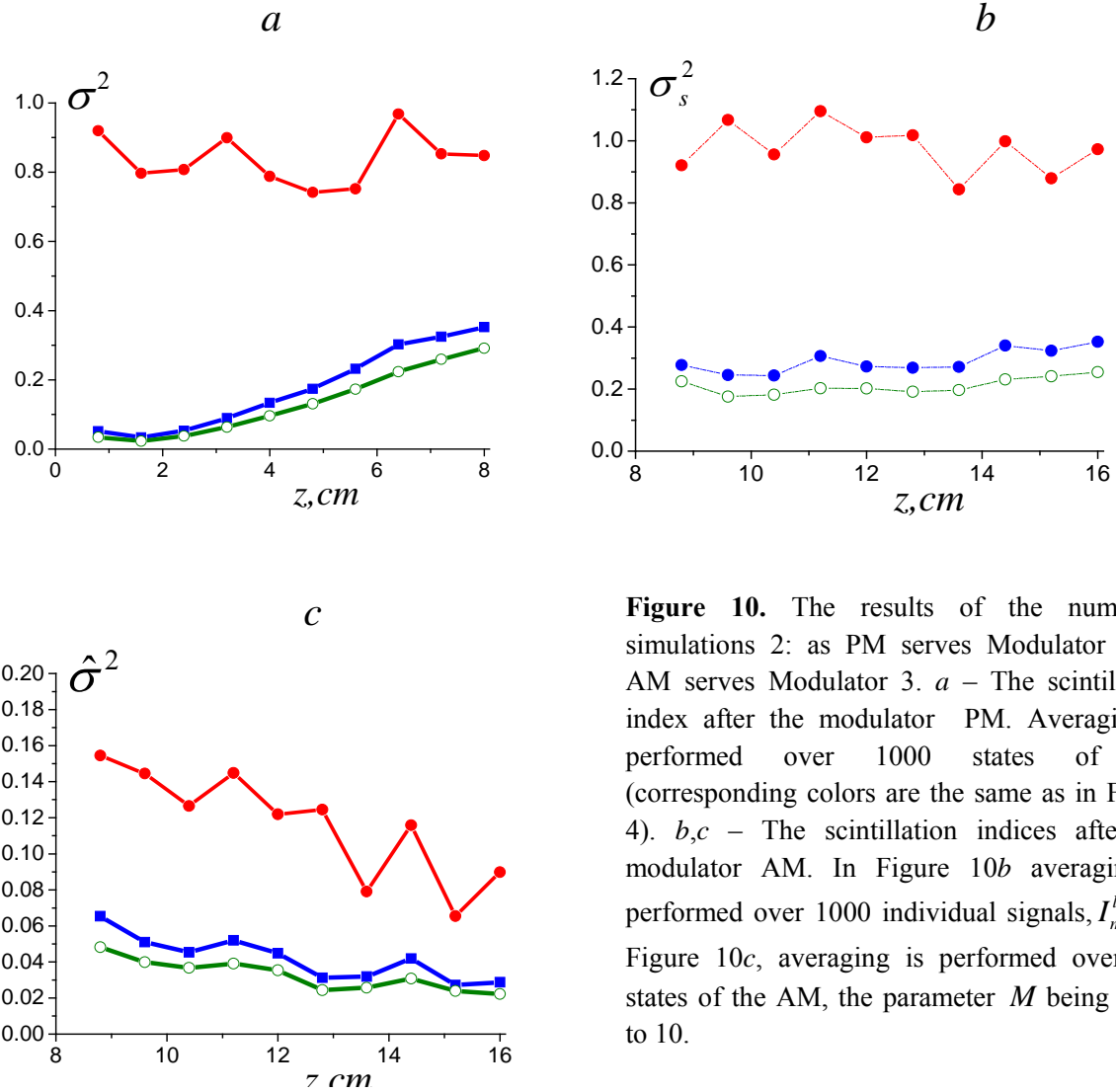

**Figure 10.** The results of the numerical simulations 2: as PM serves Modulator 1, as AM serves Modulator 3. *a* – The scintillation index after the modulator PM. Averaging is performed over 1000 states of PM (corresponding colors are the same as in Figure 4). *b,c* – The scintillation indices after the modulator AM. In Figure 10*b* averaging is performed over 1000 individual signals, $I_m^l$. In Figure 10*c*, averaging is performed over 100 states of the AM, the parameter *M* being equal to 10.

*Numerical simulations 3.* In this Section, we study the dependence of scintillation reduction on the strength of the atmospheric turbulence. For the AM, we use Modulator 2. According to the property (13), the results of Fig. 4, Fig. 9*a*, and Fig.10*a* indicate that Modulator 2 simulates the atmosphere corresponding to the highest scintillation index, $\sigma^2$, compared with Modulators 1 and 3. Modulator 3 with the lowest modulation index, $A_0 = 0.01$, is used as the PM for the PCB preparation. Despite the small modulation index, the results obtained demonstrate the strongest efficiency of the PCB (see Fig.11*b*), which can be interpreted as follows. When the PCB prepared by the modulator with a low modulation index (relatively narrow PCB, $R(d)/r_0 \approx 3$) passes the strong AM, small-scale speckle structures are formed. (Compare Figures 11*a* and 8*b*.) This phenomenon results in the highest reduction of the scintillation index for the signal integration areas R2 and R3, by factor ~30 (Compare Figures 11*b* and 9*a*). Note that the efficiency of averaging over 10 states of the PM is not very pronounced. The reduction of the scintillation index $\hat{\sigma}^2$ relative to $\sigma_s^2$ is by factor ~3 instead of 10, as found in the numerical experiments 1 and 2. The reason for this is a partial correlation between individual fragments of the PCB at $z \sim d$ (See Figure 7*a,b*) due to propagation effects. It is to be noted that the small-scale beam fragmentation (See Figures 11*a*) is the main factor leading to the efficiency of the PCB: $\sigma^2(in\ Fig.9a)/\sigma_s^2(in\ Fig.11b) \approx 10$.

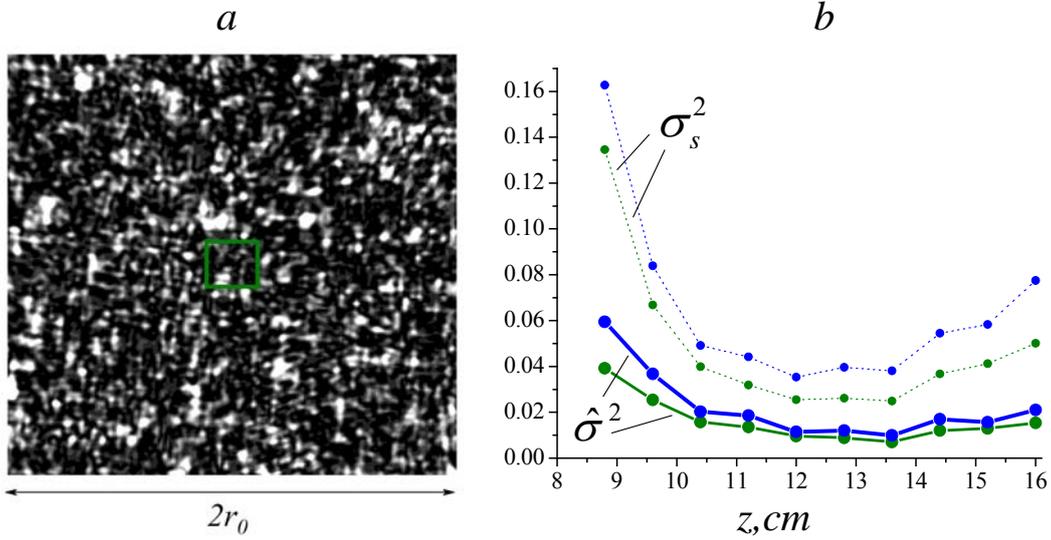

Figure 11. The results of the numerical simulations 3: for the PM we used Modulator 3, for the AM – Modulator 2. *a* – The example of random realizations, $I_m^l(x, y, z = 12 cm)$. The green square in the center of Figure 11*a* represents the integration area R3. *b* – The scintillation indices for the integration areas, R2 and R3. $\sigma_s^2$ is the result of averaging over 1000 random signals, $I_m^l$. $\hat{\sigma}^2$ is calculated over a statistical ensemble of 100 AM states; $M = 10$.

*Summary*: The main conclusions of our numerical experiments are:

i) Using a PCB in combination with averaging over random phase distributions of the phase modulator (PM), which forms the PCB, results in a significant reduction in the scintillation index, compared with an initially coherent beam.

ii) The efficiency of the scintillation index reduction depends on the size of speckles generated by the PM; the smaller the speckle size, the stronger the scintillation reduction. A PM with a higher phase modulation index provides smaller scale speckle structure, and a greater reduction of scintillations.

iii) The scintillation index reduction efficiency depends on the strength of the atmospheric turbulence. Stronger turbulence causes a more dramatic reduction (up to 30 times) of the scintillation index.

**III. Experimental results**

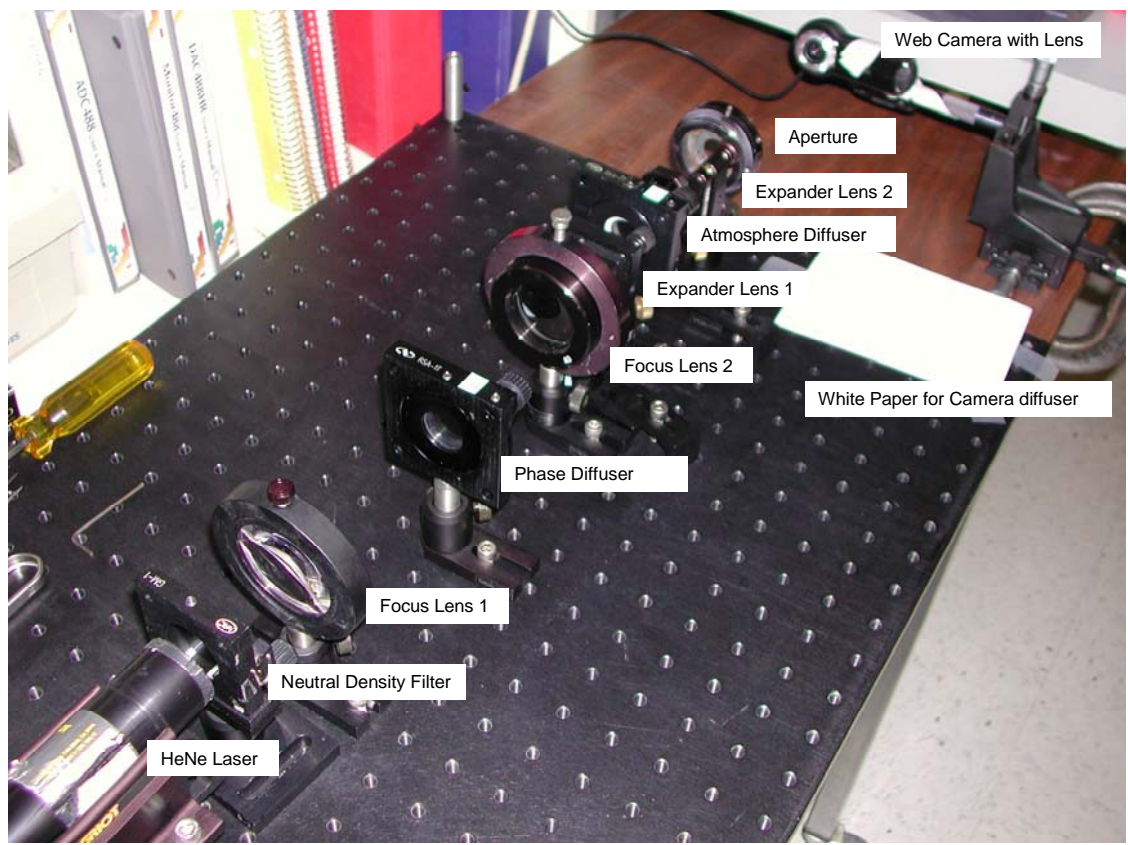

**Figure 12.** Our experimental setup.

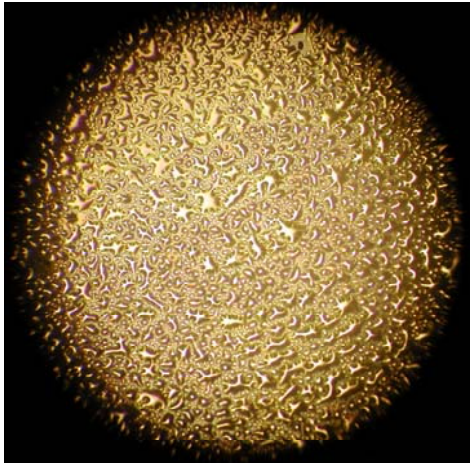

**Figure 13.** The microstructure of the phase diffuser observed through the microscope.

The experimental setup is shown in Figure 12. The coherent beam of the HeNe laser has the following specifications. The wavelength is $\lambda = 633 nm$, the output power is 0.5 mW, the mode spacing is 1078 MHz, the beam diameter is 0.47 mm, the far-field divergence is 1.7 mrad, the polarization randomness is 15% for multimode operation. The phase diffuser, which forms the PCB, has a $1°$ diffusion angle, the atmospheric phase diffuser has $0.5°$ diffuser angle, the lenslets have size in the range from 10 $\mu m$ to 50 $\mu m$ (see Figure 13). The diameters of the phase diffusers are 25 mm. The distance between the diffusers is 8 cm. The distance between the AM diffuser and the CCD plane is 8 cm. The CCD camera size is approximately 3mm by 4mm, with pixel size 2.8 $\mu$m×2.8 $\mu$m. In the experiments, the signal illuminated the CCD areas of $S = 15 \times 15$, $20 \times 20$, $40 \times 40$ and $100 \times 100$ pixels. Typical intensity distributions of the coherent beam passed through the diffusers are shown in Figures 14*a,b*.

To simulate random fluctuations in the atmosphere, the position of the AM was changed randomly in the *x-y* plane; in addition the AM was rotated randomly around the *z*-axis. The integral intensity for each integrating area was measured for 1010 positions. The scintillation index was calculated at the distance of 16 cm. Then the same procedure was used for the PM, which forms the PCB. The average signal was calculated for 10 random positions of the PM and a fixed position of the AM:

$$\hat{I}^l(S) = \frac{1}{10}\sum_{m=1}^{10} I_m^l(S), \; l = 1, 2, ..., 101 \;. \tag{15}$$

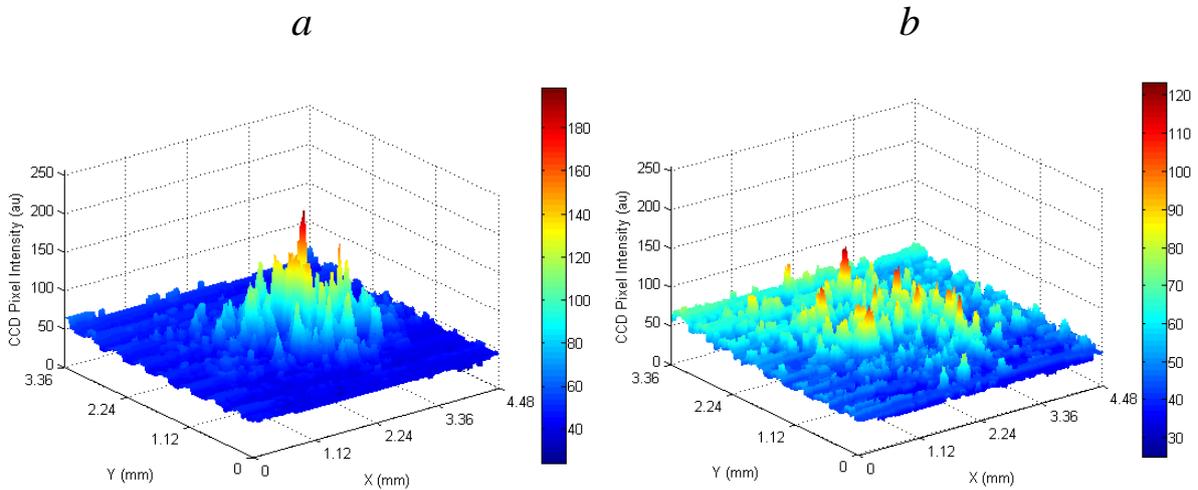

**Figure 14.** The typical intensity distribution of the beam after passing the AM (a) and PM (b) at the distance from diffusers of 8 cm. (Compare these data with the result of the numerical simulations in Figure 3.)

Table 1. Comparison of the experimental results with numerical simulations

| Receiver | $S(pixel^2)$ | Experiment | | | Numerical simulations 2 | | |
|---|---|---|---|---|---|---|---|
| | | $\sigma^2$ | $\hat{\sigma}^2$ | $\sigma^2/\hat{\sigma}^2$ | $\sigma^2$ | $\hat{\sigma}^2$ | $\sigma^2/\hat{\sigma}^2$ |
| $R1$ | $15\times 15$ | 0.313 | 0.0197 | 16 | 0.34 | 0.028 | 12 |
| $R2$ | $20\times 20$ | 0.288 | 0.0173 | 16.7 | 0.32 | 0.022 | 14.6 |
| $R3$ | $40\times 40$ | 0.182 | 0.0107 | 17 | - | - | - |
| $R4$ | $100\times 100$ | 0.056 | 0.0062 | 9 | - | - | - |

Note that the control parameters of the phase modulators, $\gamma, A_0, \Lambda_0$ used in the numerical simulations for Modulators 1 and 3 are close to those in the experiment for the PM and AM, correspondingly. Thus, the case of the numerical simulations 2 is in good agreement with experiment. The difference is related to the size distribution of lenslets. The theoretical distribution around 40 $\mu m$ is narrower than the distribution in the experiment. Including a lenslet size of 10 $\mu m$ requires much more calculation time.

## IV. Summary

We have demonstrated experimentally and numerically that the application of a PCB in combination with time averaging leads to a significant reduction in the scintillation index. We used a simplified experimental approach, in which a phase diffuser simulates the atmospheric turbulence. Although this approach has obvious limitations, it allows one to optimize the PCBs to achieve increased scintillation reduction. The roles of the speckle size, amplitude of phase modulation, and strength of the atmospheric turbulence were determined. We obtained good agreement between our numerical simulations and our experimental results. This study provides essential information for future applications of our PCB-based methods for scintillation reduction in physical atmospheres.

## V. ACKNOWLEDGMENT

We thank A.A. Chumak for useful discussions. This work was carried out under the auspices of the National Nuclear Security Administration of the U.S. Department of Energy at Los Alamos National Laboratory under Contract No. DE-AC52-06NA25396. GPB, BMC, VNG, DCL thank ONR for support.